\documentclass[12pt,preprint]{aastex}

\begin{document}

\title{The intermediate-age globular cluster NGC 1783 in the Large Magellanic Cloud}

\author{Alessio Mucciarelli}
\affil{Dipartimento di Astronomia, Universit\`a 
degli Studi di Bologna, Via Ranzani, 1 - 40127
Bologna, ITALY}
\email{alessio.mucciarelli@studio.unibo.it}

\author{Livia Origlia}
\affil{INAF - Osservatorio Astronomico di Bologna, Via Ranzani, 1 - 40127
Bologna, ITALY}
\email{livia.origlia@oabo.inaf.it}

\author{ Francesco R. Ferraro}
\affil{Dipartimento di Astronomia, Universit\`a 
degli Studi di Bologna, Via Ranzani, 1 - 40127
Bologna, ITALY}
\email{francesco.ferraro3@unibo.it}


\begin{abstract}  

We present {\sl Hubble Space Telescope} ACS deep photometry
of the intermediate-age 
globular cluster NGC 1783 in the Large Magellanic Cloud. 
By using this photometric dataset, we 
have determined the degree of ellipticity 
of the cluster ($\epsilon$=0.14$\pm$0.03) and  
the  radial density profile. This profile is well reproduced by a standard 
King model with an extended core ($r_c$=24.5'') and a low concentration 
(c=1.16), indicating that the cluster has not experienced the collapse of the core.

We also derived the cluster age, by using 
the {\sl Pisa Evolutionary Library} (PEL)  isochrones, with 
three different amount of {\sl overshooting} 
(namely, $\Lambda_{os}$=0.0, 0.10 and 0.25). 
From the comparison of the  observed
Color-Magnitude Diagram (CMD) and 
Main Sequence (MS) Luminosity Function (LF) with 
the theoretical isochrones and LFs,
we find that only models with the inclusion of some {\sl overshooting} ($\Lambda_{os}$=0.10-0.25) 
are able to reproduce the observables.
By using 
the magnitude difference  $\delta V_{SGB}^{He-Cl}=0.90$ between 
the mean level of the He-clump and the
flat region of the SGB, we derive an age $\tau$=1.4$\pm$0.2 Gyr.

\end{abstract}  
 
\keywords{Magellanic Clouds --- globular clusters: individual (NGC~1783) ---
techniques: photometry}   

\section{Introduction}   
\label{intro}

Stellar clusters are key-tracers of stellar populations in different  galactic
environments.  In particular, populous clusters in the Large Magellanic Cloud
(LMC) cover a wide range of ages (from a few Myr up to 13 Gyr) which has no
counterpart in  our Galaxy. Hence, the study of this system allows to extend our
empirical  knowledge of stellar populations in a mass regime which 
can be poorly explored in our Galaxy.\\ 
The LMC clusters can be grouped in three
main age families, namely:  the young population with ages $\le$200  Myr
\citep{valle94,testa99}, the intermediate population in the 200 Myr $<$ age $<$ 
3-4 Gyr range \citep{f95,broc01,gallart03,f04} and the old population, with stellar
clusters coeval to the Galactic Halo ones \citep{testa95, brocato96, olsen98,
mac04}.\\
A few
decades ago, the main integrated properties of the LMC cluster system, both in the infrared 
\citep{persson83} and in the optical \citep{ma79, ma82, swb, vdb81} spectral ranges  
have been investigated. These studies also provided the only existent homogeneous age-scale, 
based on the so-called {\sl s-parameter}, as defined by \citet{ef85}. This parameter is an empirical 
quantity related to the position of the clusters in the (U-B)-(B-V) color-color diagram.
Clearly, this method presents many uncertainties, namely the
foreground/background  contamination and the possible statistical fluctuations due
to bright stars. 

The advent of 8-meter class ground-based
telescopes and the superior performances of the  Hubble Space Telescope (HST) provide 
sufficient resolution 
to properly study   these clusters even in their innermost crowded regions. 
Accurate ages can be determined from the Main
Sequence (MS) Turn-Off (TO) measurements  \citep[see e.g. the recent works
by][]{mac06,kerber,m07},  once updated theoretical evolutionary models are
adopted and precise estimates  of the global metallicity \citep{salaris} be
available. Indeed, the stellar clock is extremely sensitive  to the chemical
composition and detailed abundances of iron and $\alpha$-elements  from
high-resolution spectroscopy are mandatory for this purpose.

A few years ago, we started a long term project aimed  at determining homogeneous ages and
metallicities for a representative sample  of template LMC clusters, by combining
high-resolution photometry and spectroscopy.  The first cluster analyzed so far is
NGC 1978: an accurate metallicity of [Fe/H]=-0.37$\pm$0.07 dex \citep{f06} and an
age of $\tau$=1.9$\pm$0.1 Gyr \citep[][hereafter Paper I]{m07} have been obtained.  In
this paper we present the results for NGC~1783, another populous intermediate-age cluster.
Sect.~\ref{obs} describes the cluster Color-Magnitude Diagram (CMD) and
its main  evolutionary features. Sect.~\ref{parsec} describes its structural parameters, 
while 
Sect.~\ref{agesec} discusses its age determination. 
In Sect.~\ref{disc} we draw our conclusions.

\section{Observations and data analysis}
\label{obs}

\begin{figure}[h]
\plotone{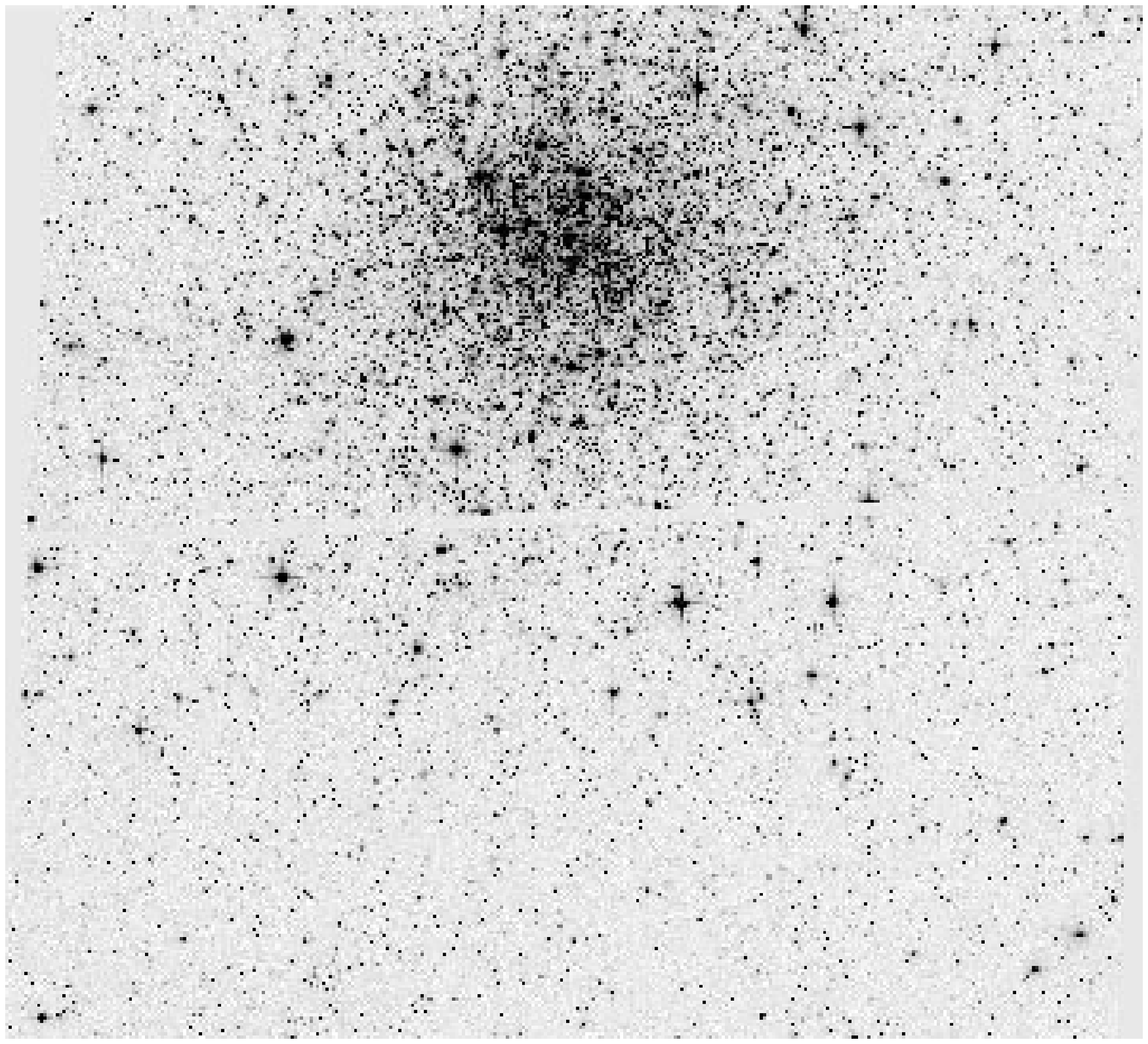}
\caption{ACS/WFC F814W image of the LMC cluster NGC 1783, both two chips.}
\label{fits}
\end{figure}

The results presented in this paper are based on a set of images obtained 
with the Advanced Camera for Survey (ACS) Wide Field Channel (WFC) that 
provides a field of view of $\approx$200'' $\times$200'' with a
plate scale of 0.05 arcsec/pixel. All the images have been retrieved 
from the ESO/ST-ECF Science Archive (Proposal ID 9891, Cycle 12), 
through the F555W and F814W filters, with
exposure times of 250 and 170 sec, respectively. 
The first chip of the ACS-WFC 
is centered on the cluster center. 
Fig.~\ref{fits} shows the
the F814W image of the cluster in both the ACS chips.\\
The photometric reduction was carried out with the  
{\it DAOPHOT-II} package \citep{stet} by using the 
Point Spread Function (PSF) fitting method. 
The final photometric catalog includes almost 40,000 stars, 
and it has been calibrated in the VEGAMAG photometric system using the 
prescriptions of \citet{bedin} and astrometrized on the 
Two-Micron All-Sky Survey (2MASS) photometric system,
by cross-correlating the ACS@HST catalog with  the infrared catalog
presented by  \citet{m06}.

\subsection{The CMD overall characteristics}
\label{cmdsec}

\begin{figure}[h]
\plotone{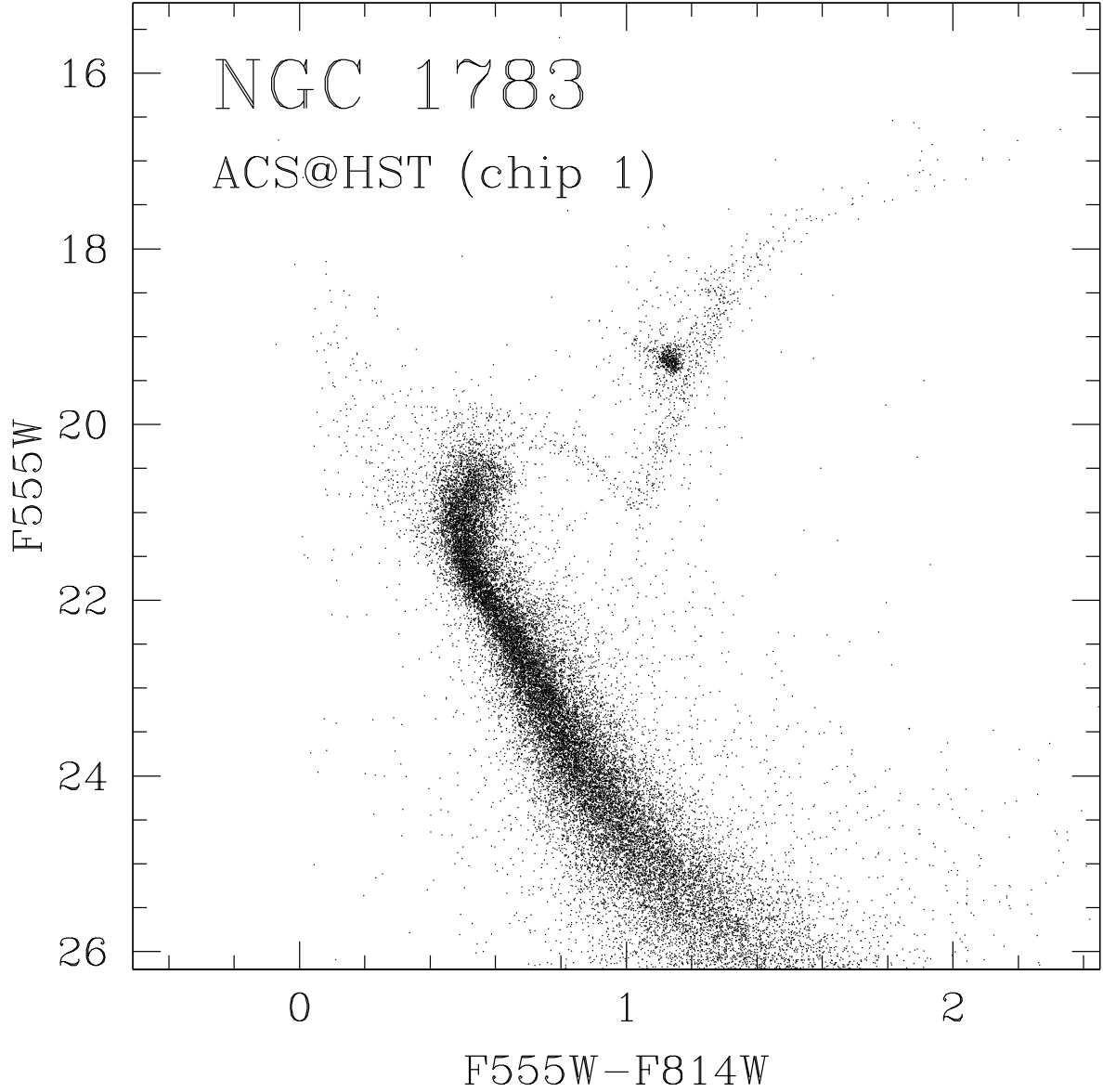}
\caption{(F555W, F555W-F814W) CMD of the LMC cluster NGC 1783,
 obtained with ACS@HST (only  stars lying into the chip  containing the cluster
 core have been plotted).}
\label{cmd}
\end{figure}

Fig.~\ref{cmd} shows the observed CMD
using only the ACS chip sampling the cluster core.  
The useful magnitude range is $\rm 17.6\le F555W \le 26$. 
Indeed, we note that the brightest stars 
at F555W$<$17.6, could 
be in the non-linear regime of the CCD or saturated in their central pixels, 
making the 
corresponding magnitudes and colors somewhat uncertain.\\
The main features of the observed CMD can be summarized as follows:\\ 
(1) The MS extends over more than 6 magnitudes in the 
F555W band and the TO point is located at F555W$\approx$21.2 
(the identification of the TO magnitude was done by means of a 
parabolic fit of this region). 
The TO region shows a mild spread in color;\\ 
(2) the slope change of the MS is at  F555W$\approx$22.2 and 
flags the transition between radiative and convective core stellar structures;\\ 
(3) the Sub Giant Branch (SGB)  is a poorly populated sequence, 
with a typical F555W$\approx$20.5 magnitude.
We note that the blue edge of this sequence is not well-defined;\\ 
(4) the Red Giant Branch (RGB) is well populated and it extends 
over $\approx$5 magnitudes;\\  
(5) the Helium-Clump is located at F555W$\approx$19.25 and
(F555W-F814W)$\approx$1.15; \\ 
(6) the Asymptotic Giant Branch (AGB) Clump 
(corresponding to the base of the AGB sequence)                                   
is visible at F555W$\approx$18.4. 
       
\begin{figure}[h]
\plotone{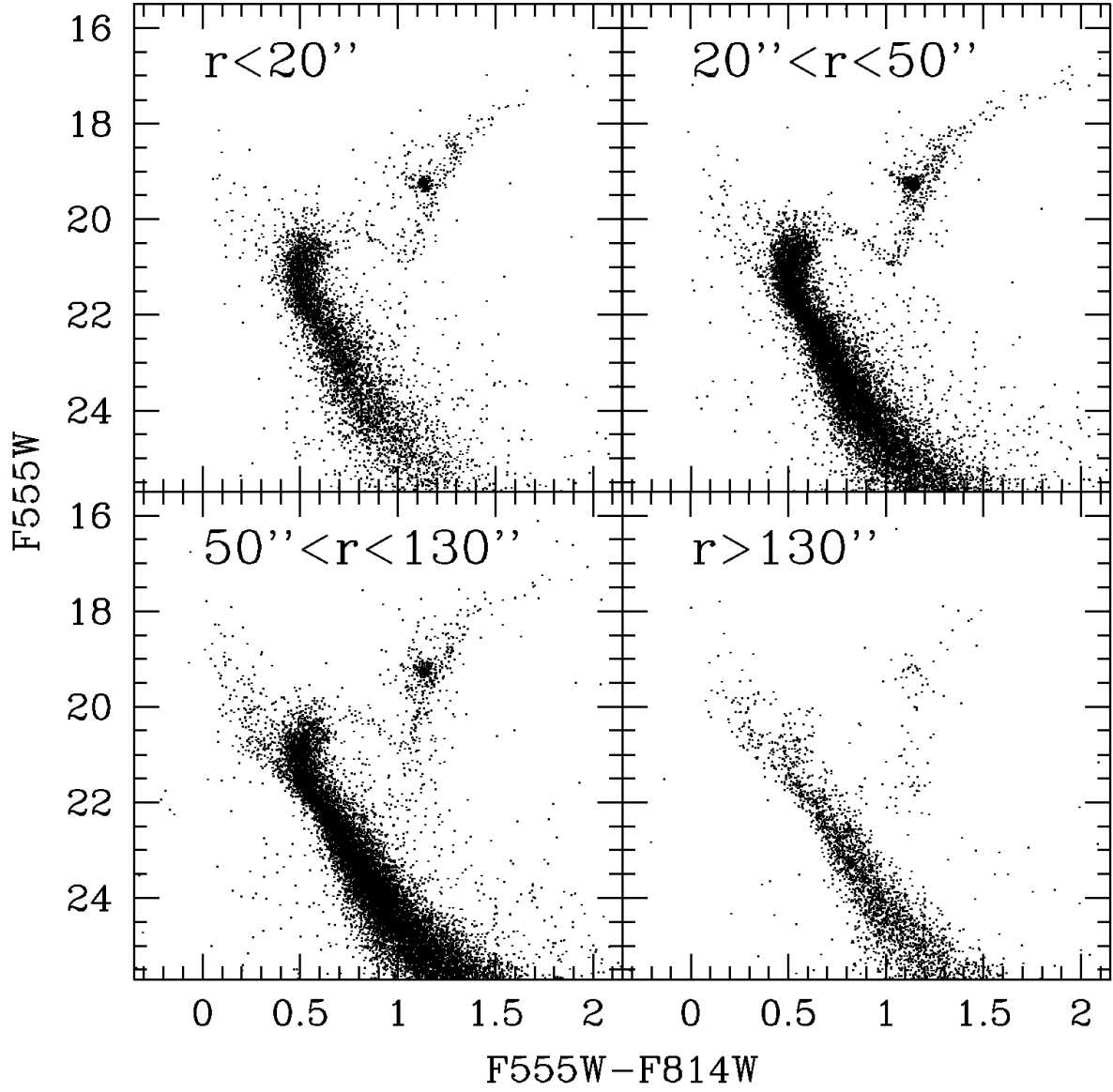}
\caption{Radial (F555W ,F555W-F814W) CMD of NGC 1783 at increasing distances from the cluster
center. }
\label{radcmd}
\end{figure}

Fig.~\ref{radcmd}  shows the radial 
CMDs by using   the entire sample of stars detected in the ACS FoV. 
The bulk of the cluster population lies in the central 2 arcmin 
(by radius); at r$>$130'' the SGB,   
RGB and He-Clump are barely detectable,  while the brightest portion of the
cluster MS is still visible.

The mild color broadening of the TO region deserves a brief discussion. 
Recently, \citet{ber} found a color dispersion in the brightest
portion of the MS of NGC 2173, while \citet{mac07} found 
a bifurcation of the bright MS region of NGC 1846, and interpreted it
as a double TO. 
These two observational evidences suggest the possible existence 
of an age-dispersion in these stellar clusters.
In order to check whether the broadening of the TO region in NGC 1783 can be 
ascribed to a possible age-dispersion as well, 
we calculated the color distribution of the MS stars in the 20.5$<$F555W$<$21.1 magnitude range. 
The color distribution turns out to be
roughly Gaussian with $\sigma_{F555W-F814W}\approx$0.05, which is fully
consistent with the observational errors ($\sigma_{F555W}\sim\sigma_{F814W}\approx$0.03,
implying a color uncertainty $\sigma_{F555W-F814W}\approx$0.04). Similar results         
are obtained by computing the color distribution in the radial CMDs of Fig.~\ref{radcmd}.
Thus, we can conclude that the spread in color of the TO region in NGC 1783 can
be explained in terms of photometric errors and there is not any 
evidence of an age-dispersion.

\subsection{Completeness}
\label{compsec}

\begin{figure}[h]
\plotone{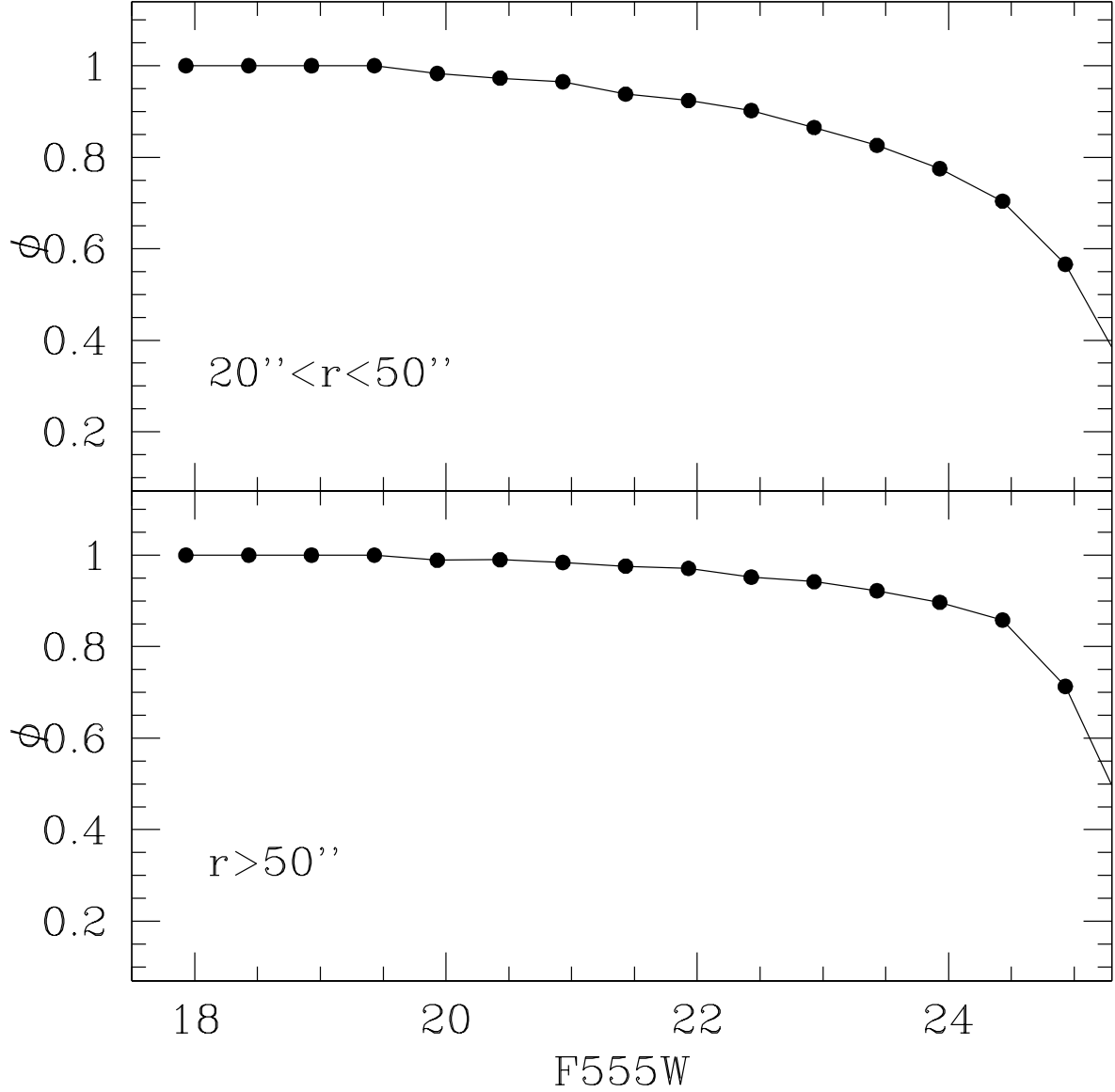}
\caption{Completeness curves computed in two radial sub-regions of NGC 1783.
 The black points indicate the value of the $\phi=\frac{N_{rec}}{N_{sim}}$ 
 parameter calculated for each 
0.5 magnitude bin.  }
\label{comp}
\end{figure}

In order to quantify the degree of completeness of the final  photometric catalog,
we  used the well-know artificial star technique \citep{mateo}, and we simulated a
population of stars in the same magnitude range  covered by the observed CMD
(excluding stars brighter than F555W=17.6, corresponding to the saturation level) and
with a (F555W-F814W)$\sim$0.8  mean color. The artificial stars have been added to the
original images and the entire data reduction procedure has been repeated using the
{\sl enriched} images. The number of artificial stars simulated in each run ($\sim$
2,000)  is always a small percentage ($\sim$5\%) of the detected stars, hence they do 
not alter the original crowding conditions. A total of $\sim$250 runs  were performed
and more than 500,000 stars have been simulated. 
We  have excluded from our analysis the very inner
region of the cluster (r$<$20''), where the crowding  conditions are prohibitive. 
Fig.~\ref{comp} shows the completeness factor $\phi=\frac{N_{rec}}{N_{sim}}$, defined as the
fraction of  recovered stars over the  total simulated ones, as a 
function of the F555W magnitude in two different radial regions, namely between 20''
and 50'' and at  r$>$50'' from the cluster center, respectively.
In the inner region the sample is $>$90\% complete down to F555W$\approx$22.5, 
while in the outer region is  $>$90\% complete down to F555W$\approx$24.

\section{Ellipticity and structural parameters}
\label{parsec}
       
The knowledge of the position of each star over the entire 
extension of the cluster (and in particular in the innermost region) 
allows to compute the center of gravity ($C_{grav}$) 
with high precision. In doing this, we applied the procedure 
described in \citet{mon95}, averaging the $\alpha$ and $\delta$ 
coordinates of the detected stars with F555W$<$22, 
in order to minimize the effects of 
incompleteness.
The $C_{grav}$ of 
the cluster turns out to be located at $\alpha$=$4^h$ $59^m$ $09^s$.78
and $\delta$=-$65^{\circ}$ 59' 17''.82. 
This finding is  in good agreement with our previous 
determination   based on 
near-IR photometry \citep{m06}.

\begin{figure}[h]
\plotone{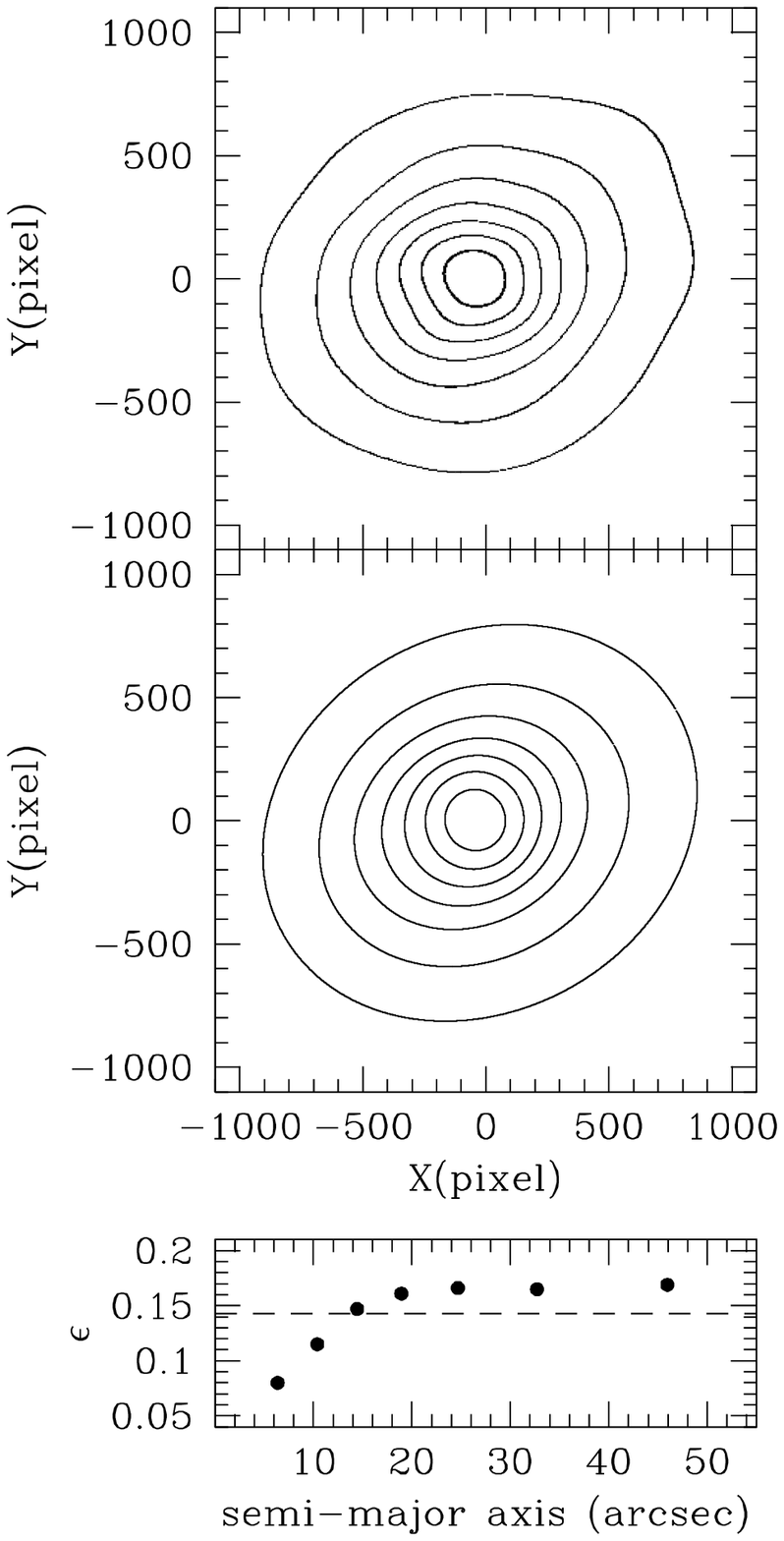}
\caption{{\it Upper panel}: the map of NGC 1783 with the isodensity contours;
 {\it central panel}: 
the best fit ellipses  to the isodensity contours; {\it lower panel}: ellipticity of 
the best fit ellipses as a function of the semi-major axis in arcsec. The 
horizontal dashed line indicates the mean value.}
\label{isoden}
\end{figure}

We also used the ACS photometry of NGC 1783 to derive new estimates  for the cluster
ellipticity and structural parameters.  The isodensity curves are computed 
with an adaptive kernel technique, accordingly to the prescription of
\citet{fuk}. We used all the stars in the first chip with F555W$<$22  in order to
minimize   incompleteness effects and we fit the isodensity curves  with ellipses. 
Fig.\ref{isoden} shows the cluster map with the isodensity contours (upper panel), the
corresponding best fit ellipses  (central panel) and their ellipticity as a function
of the semi-major axis in arcsec (lower panel). The ellipticity $\epsilon$
(defined as $\epsilon$=1-(b/a), where a and b are the major and minor axis of the
ellipse, respectively)  turns out to be 0.14$\pm$0.03.  This value results slightly
lower than the previous determinations of \citet{geisler}  that found an average
ellipticity of $\epsilon$=0.19. 

\begin{figure}[h]
\plotone{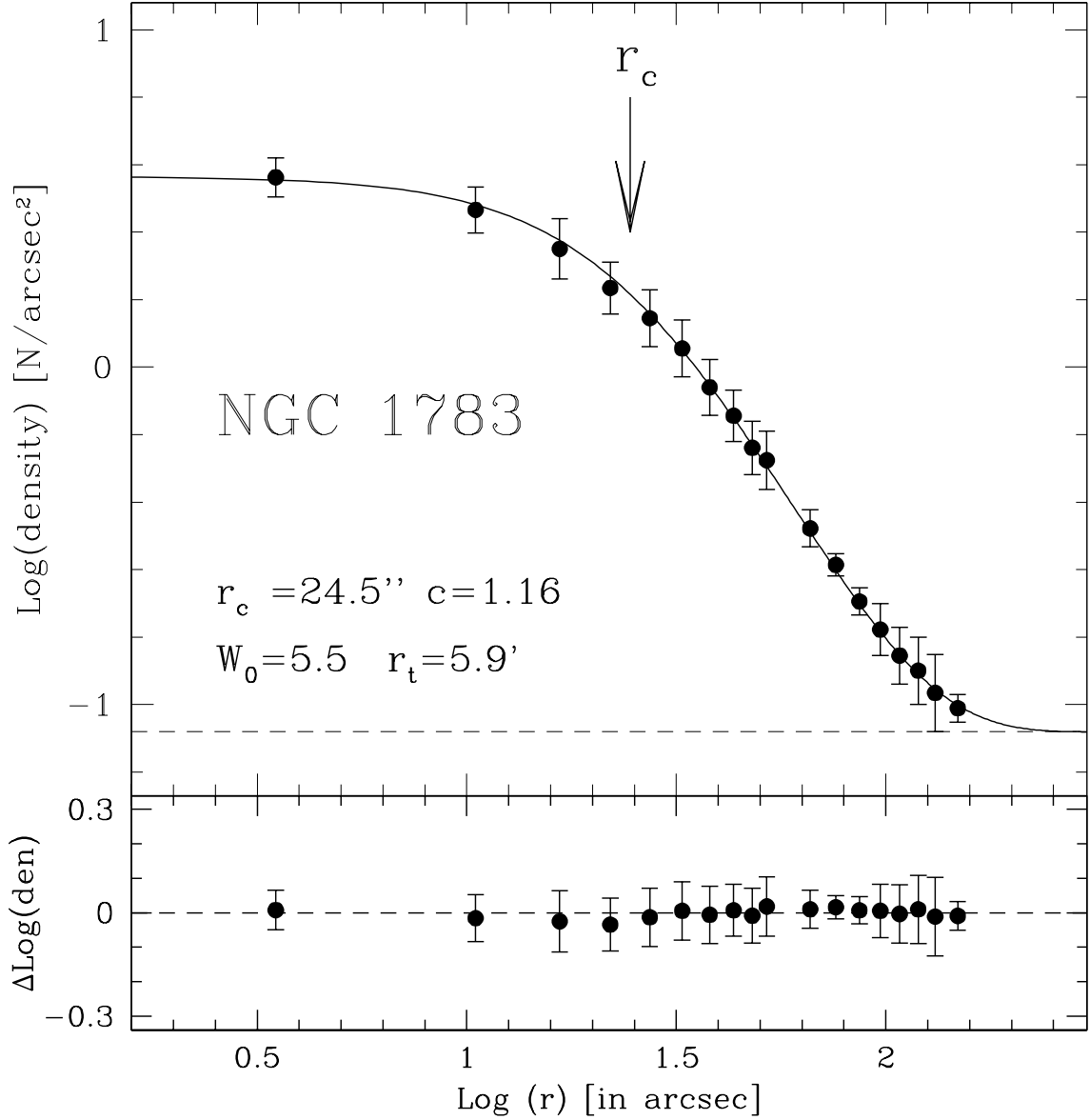}
\caption{{\it Upper panel}: observed radial density profile for the cluster NGC 1783. 
The solid line is the best fit King model, with $r_c$=24.5'' and c=1.16. 
The {\it horizontal dashed line} 
indicate the background level. {\it Lower panel}: 
the $\chi^2$ test for the observed radial density profile and best-fit King model
(solid line). }
\label{radprof}
\end{figure}

By following the procedure already described  in previous
papers \citep[see][]{f04b},
we also compute the  projected density
profile of the cluster. 
The area sampled by the first ACS chip has been  divided
in 18 concentric annuli, each one centered on $C_{grav}$ and split in four sub-sectors. 
The number of stars lying in each sub-sector was counted and the mean  star density 
was obtained. The standard deviation was estimated from the variance among  the
sub-sectors. The radial density profile is plotted in ~Fig.\ref{radprof}. \\  We used the
\citet{sig} code in order to compute the family  of isotropic single-mass King
models.  These models are  defined by three main parameters, the central potential
$W_0$,  core radius $r_c$ and the concentration 
c=$\log{(r_t)/(r_c)}$, where $r_t$ is the tidal radius.  
Fig.~\ref{radprof} also shows the single-mass
King model that best fit the derived  density profile. The best-fit model  has been 
selected
by using a $\chi^2$ minimization (shown in the lower panel of Fig.~\ref{radprof}). 

We find
$W_0$=5.5, $r_c$=24.5'' and c=1.16 , corresponding to a tidal radius $r_t$= 5.9' 
\footnote{We underline that the structure of the profile and the corresponding
derived parameters does not change if different magnitude limits are adopted.}. 
Our estimate of $r_c$ is consistent with the one by \citet{elson92} who found 
$r_c$=20''. 
The resulting $r_t$  lies out of the field of view of ACS.  
In order to properly fit the  most external  points
of the radial profile,
the best-fit King model has been combined with a constant background level
(corresponding to a density of 350 stars $arcmin^2$), and shown as a
 horizontal dashed line 
in Fig.~\ref{radprof}. 

\section{The age of NGC 1783}
\label{agesec}

Young stellar populations (with ages $\le$300 Myr) are characterized by large
convective cores. 
Theoretical studies  \citep[see e.g. the numerical
simulations computed by][]{freytag}  suggest that  
the penetration of convective
elements into a stable region ({\sl via the Schwarzschild criterion})
can produce  non-negligible evolutionary effects.
These prediction seem to be confirmed by several works  
\citep{bm83,barmina,chiosi07} which require 
some amount of {\sl
overshooting} in the MS  star convective core , in order 
to reproduce the observed morphologies
and stellar counts of young clusters, although this issue 
is still matter of
debate \citep{testa99,broc03}. 
At variance, in older ($\ge$5-6 Gyr) stellar populations the
growth of large radiative cores  tends to erase the possible evolutionary effects 
of {\sl overshooting}. 

Intermediate-age stellar populations like those in NGC 1978 and 
NGC 1783 LMC stellar clusters represent the transition stage between 
these two regimes, and 
thus represent ideal test-bench to study the {\sl overshooting} effects.

\subsection{Basic assumptions}
\label{basic}

In Paper I we 
 performed a detailed
comparison of the observed morphology and star counts 
of NGC 1978 with different set of theoretical
models and {\sl overshooting} efficiencies.
The best agreement between observations and theoretical predictions was 
reached with the Pisa Evolutionary Library (PEL). 
\footnote{The PEL isochrones are available at the URL 
http://astro.df.unipi.it/SAA/PEL/Z0.html.} 

Hence, we have used the PEL isochrones to also determine the age of NGC 1783.
We select isochrones with Z=0.008 (corresponding to
[M/H]=-0.40 dex, as estimated by Mucciarelli et
al.(2007, in preparation) from high-resolution spectroscopy), and with three
different amount of {\sl overshooting} efficiency, namely
$\Lambda_{os}$
\footnote{The overshooting efficiency is parametrized 
using the mixing length theory
\citep{bome} with $\Lambda_{os}$=1/$H_p$ 
(where $H_{p}$ is pressure scale height) that quantifies the overshoot distance {\sl above} 
the Schwarzschild border in units of the pressure scale height.}
=0.0 for the canonical isochrones, and $\Lambda_{os}$=0.10
and  0.25, representative of mild and strong {\sl overshooting} regimes, respectively.

These theoretical isochrones have been transformed into the observational plane,
by means of suitable conversions computed with the code described
by \citet{origlia},  and convolving the model atmospheres by \citet{bcp}
with the ACS filter responses.  Guess values 
of $(m-M)_0$=18.50 \citep{alves} for the distance modulus and
E(B-V)=0.10 \citep{persson83} for reddening have been adopted.  However, 
in order to obtain
the best fit of the observed sequences we allowed  these parameters
to vary by $\le|10|$\% and $\le|40|$\% factors, respectively.

\begin{figure}[h]
\plotone{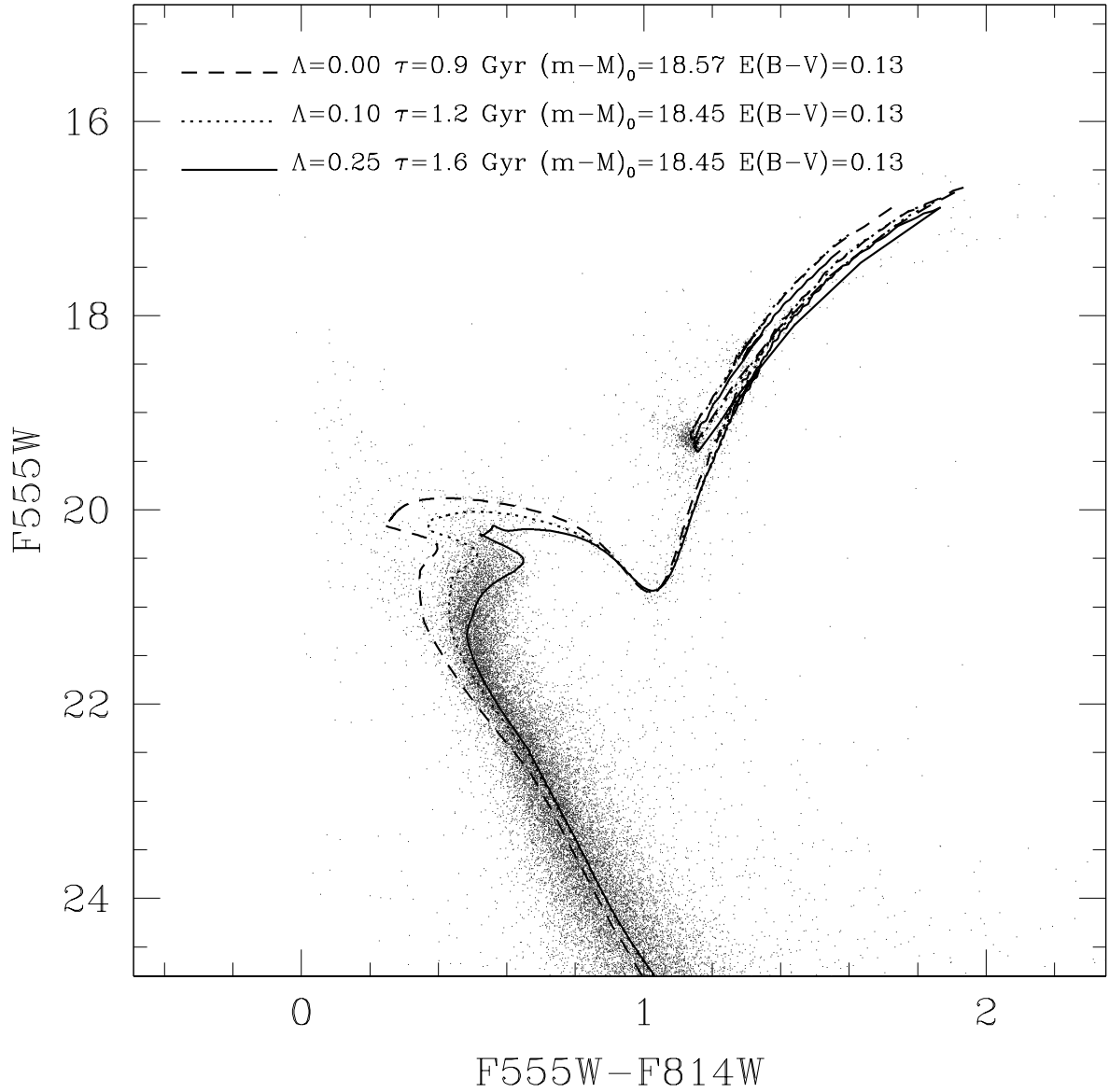}
\caption{Best-fit theoretical PEL isochrones overplotted on the  observed 
CMD of NGC 1783. Models with different assumptions of the overshooting
efficiency ($\Lambda_{os}$) are used: the best fit age,
distance modulus and reddening (see text) for each
choice of $\Lambda_{os}$ are also marked.}
\label{isoc}
\end{figure}

Fig.~\ref{isoc} shows the best-fit solutions
for the different values 
of $\Lambda_{os}$, as obtained by
matching the following features:\\ 
(1) the magnitude of the He-Clump;\\ 
(2) the magnitude difference between the He-Clump and the flat 
region of the SGB;\\ 
(3) the difference in color between the TO 
and the base of the RGB.

As can be seen, the canonical model with $\Lambda_{os}$=0.0 fit the 
observational features (1) and (2) reasonably well 
with $(m-M)_0$=18.57, E(B-V)=0.13 and 
$\tau$=0.9 Gyr,
but fails to reproduce feature (3). 

\begin{figure}[h]
\plotone{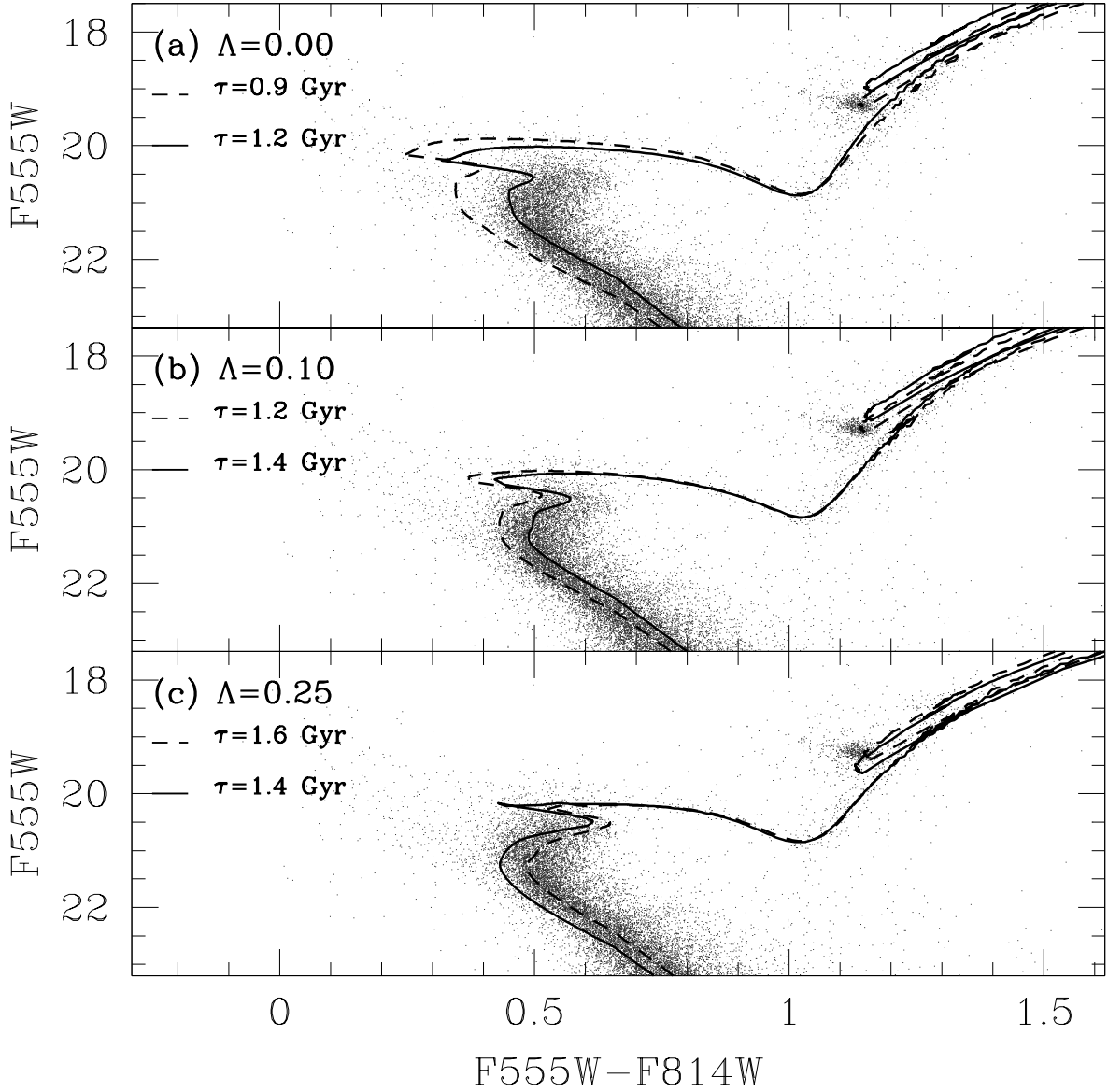}
\caption{
Portion of the CMD, as zoomed onto the TO region,
with different models overplotted.
Panel (a):  best-fit (dashed line, $\tau$=0.9 Gyr, $(m-M)_0$=18.45) and older (solid line, 
$\tau$=1.2 Gyr, $(m-M)_0$=18.16) isochrones for $\Lambda_{os}$=0.00. 
Panel (b):  best-fit (dashed line, $\tau$=1.2 Gyr, $(m-M)_0$=18.45) and older (solid line, 
$\tau$=1.4 Gyr, $(m-M)_0$=18.25) isochrones for $\Lambda_{os}$=0.10.
Panel (c):  best-fit (dashed line, $\tau$=1.6 Gyr, $(m-M)_0$=18.45) and youngerer (solid line, 
$\tau$=1.4 Gyr, $(m-M)_0$=18.66) isochrones for $\Lambda_{os}$=0.25.
}
\label{isoc2}
\end{figure}

Fig.~\ref{isoc2} (panel (a)) shows a portion of the CMD, as zoomed onto the TO region,
with the best-fit ($\tau$=0.9 Gyr) and 0.3 Gyr older ($\tau$=1.2 Gyr) isochrones.
The older isochrones better fits feature (3) but predicts a too bright
(by $\approx$0.3 magnitudes) He-clump.
Moreover, it requires a $(m-M)_0$=18.16 distance modulus, which is
definitely too short for the LMC \citep{alves}.\\
Fig.~\ref{isoc2} (panels (b) and (c)) shows
a similar comparison for the overshooting models.
For the $\Lambda_{os}$=0.10 model (panel (b)), the best-fit ($\tau$=1.2 Gyr) and 0.2 Gyr older
($\tau$=1.4 Gyr) isochrones are plotted.
As for the canonical model,
the older isochrones somewhat better fits feature (3) but predicts a too bright
(by $\approx$0.25 magnitudes) He-clump and a too short
$(m-M)_0$=18.25 distance modulus.\\
For the $\Lambda_{os}$=0.25 model (panel (c)), the best-fit ($\tau$=1.6 Gyr) and 0.2 Gyr younger
($\tau$=1.4 Gyr) isochrones are shown.
The younger isochrone sligthly better fits the SGB region but predicts a too blue MS.
Also it predicts a sligthly too faint (by $\approx$0.2 magnitudes) He-clump
and too long $(m-M)_0$=18.66 distance modulus.\\
In summary, we can conclude that canonical models,
regardless the adopted isochrone age,
do not provide an acceptable fit
to the observed CMD, while models with
$\Lambda_{os}$=0.10 and 0.25 {\sl overshooting},
E(B-V)=0.13, $(m-M)_0$=18.45
and ages between $\tau$=1.2 and $\tau=$1.6 Gyr, respectively,
reasonably well reproduce all the three diagnostics features.

\subsection{Star counts and overshooting efficiency}
\label{count}

A quantitative check to 
discriminate between the different   {\sl overshooting scenarios} is to 
perform a comparison between the observed and theoretical LFs 
of the MS stars normalized to the number of the He-clump stars, defined as 
$$\Phi_{norm}=\lg{\frac{\sum_{i}{N_{MS}}}{N_{He-Cl}}}.$$ 
Such a normalized LF is a powerful indicator of the relative timescales of the H and He burning 
phases. 
The observed $\Phi_{norm}$ is obtained by counting the number of MS stars 
($N_{MS}$) in each 0.5 magnitude bin, after the  
correction for incompleteness and field contamination, and normalized 
to the total number of He-Clump stars. 
The innermost  region of the cluster  ($r<20''$, see Fig.~\ref{radcmd}) has been
excluded from this analysis 
because of its prohibitive crowding.  
Formal errors for the observed $\Phi_{norm}$ in each magnitude bin 
are computed under the assumption that star counts 
follow the Poisson statistics, 
by using the following formula:
$$\sigma_{\Phi_{norm}}=\frac{\sqrt{{\Phi_{norm}}^{2}\cdot\sigma_{N_{He-Cl}}^{2}+\sigma_{N_{MS}}^{2}}}{N_{He-Cl}}.$$ 

Since the ACS field of view is not large enough to properly sample 
the field population around NGC 1783, 
we used the most external region ($r>150''$) of the
decontamination field for NGC1978 (Paper I).
Indeed, these two clusters are close enough for the purpose of decontamination and 
their field RGB sequences are well-overlapped.

\begin{figure}[h]
\plotone{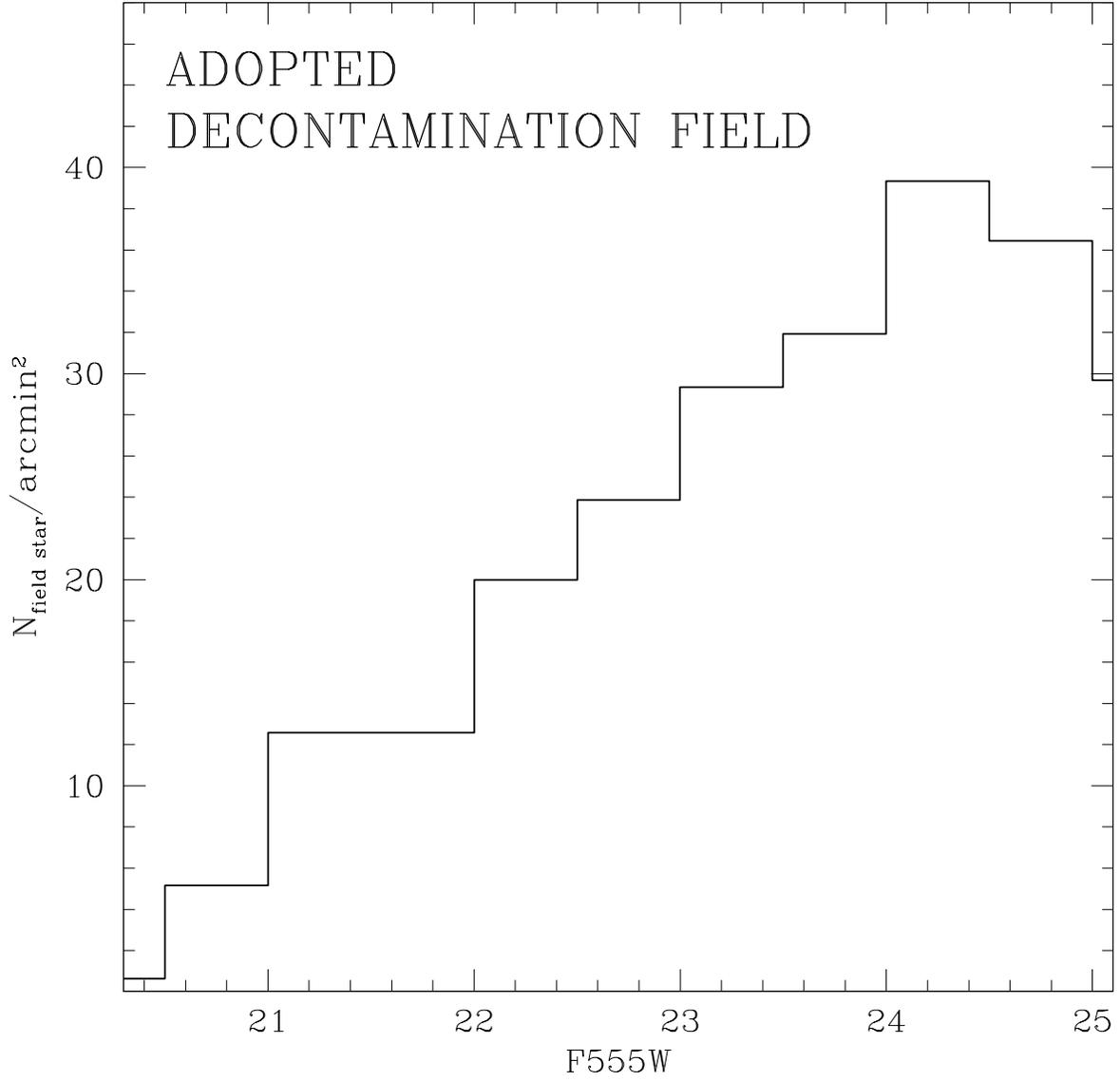}
\caption{Histogram of the number of MS stars per $arcmin^{2}$
at r$>$150'' from the center of NGC 1978.}
\label{dec}
\end{figure}

Fig.\ref{dec} shows the histogram of the number of MS stars per $arcmin^{2}$
at r$>$150''  from the center of
NGC 1978. 
The number of MS and He-Clump stars in this field have been 
subtracted from the NGC 1783 cluster stellar counts,
after the normalization for the sampled area.

Hence, the total number of stars in 
each magnitude bin is given by:
$$N_{corr}=\frac{N_{obs}}{\phi}-N_{field}.$$ 

In order to compute the theoretical $\Phi_{norm}$ predicted by the PEL models, we have 
adopted the well-know technique of synthetic diagrams. By using the 
best-fit models described above, we randomly distributed the stars 
along the isochrone accordingly to a Salpeter initial mass function.
An artificial 
dispersion has been added  in order to simulate the photometric errors. 
For each model, 200 
synthetic diagrams are computed by using Montecarlo simulations, and 
the corresponding $\Phi_{norm}$ are extracted
and averaged together. \\

\begin{figure}[h]
\plotone{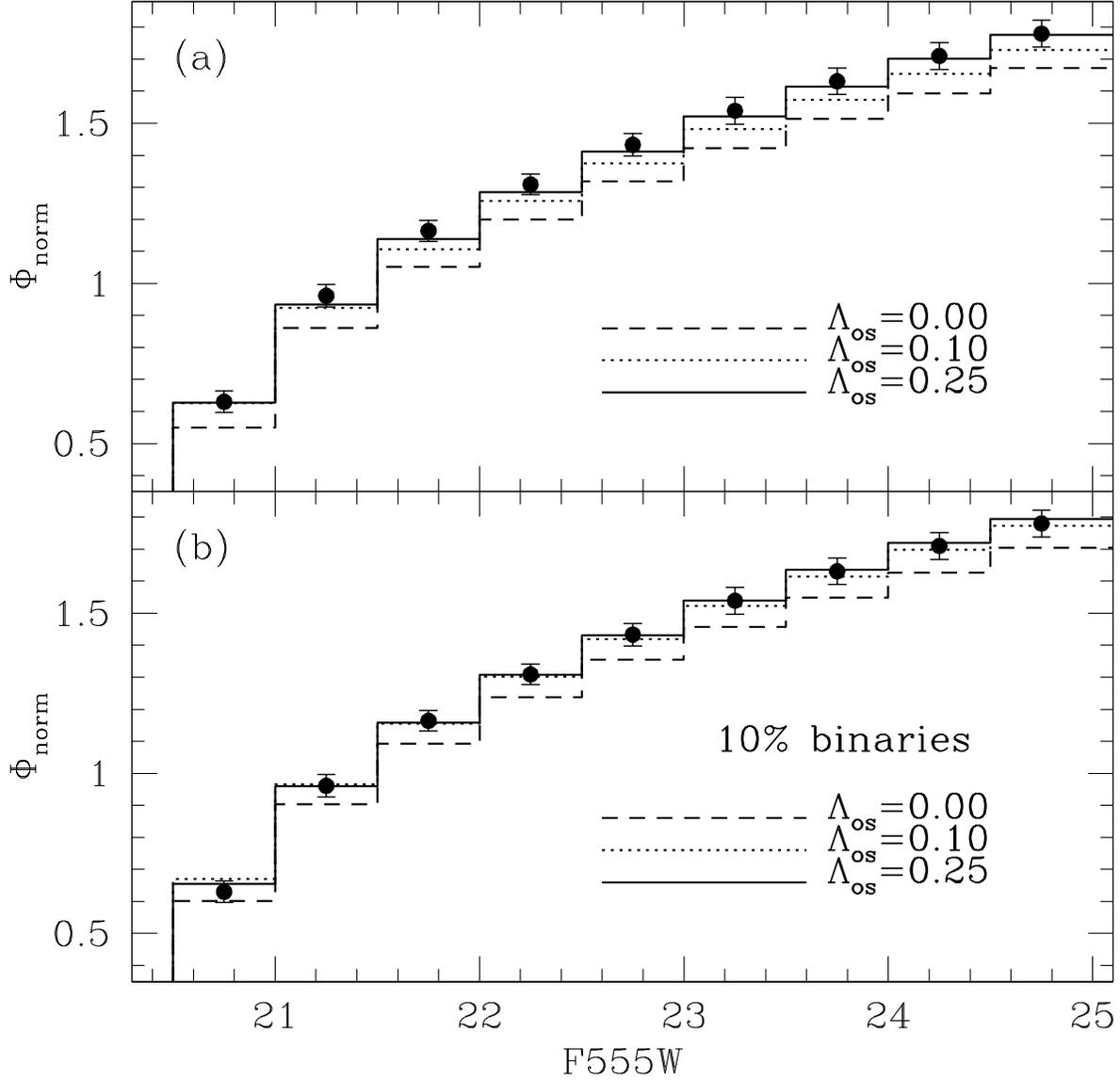}
\caption{Panel {\sl (a)}: integrated LF of the 
MS stars normalized to the number of 
He-Clump stars: {\it black points} indicate the observed LF 
and their size 
correspond to their typical uncertainty. 
The three line are the theoretical 
LFs computed by adopting $\Lambda_{os}$=0.0 (dashed), 0.10 (dotted) and 0.25 
(continuous). Panel {\sl (b)}: 
same as panel {\sl (a)}, but adding 
a 10\% binary fraction in the computation of the theoretical LFs.}
\label{lumfun}
\end{figure}

Fig.~\ref{lumfun} (panel {\sl (a)}) shows the observed LF (black points)
compared with the theoretical expectations, 
computed by using the three different {\sl overshooting} models.
Clearly, the $\Lambda_{os}$=0.0 model 
predicts a $\Phi_{norm}$ value $\sim$10-15\% lower than the observed one; 
the  $\Lambda_{os}$=0.10 and 0.25 models marginally ($<$5\%) 
underestimate the observed value of $\Phi_{norm}$.
This small offset can be easily accounted for by adding a binary population 
in the synthetic LF. To do this, we assumed that a
given fraction {\sl $f_b$} of the simulated stars be  
the primary star of a binary system. 
The mass of the
primary is randomly extracted, while 
the mass of the secondary star 
is assigned by adopting the mass ratio {\sl q}, 
between the secondary and primary star.
The magnitude of the binary system is given by 
$M_{F555W}^{Binary}=-2.5\cdot\log{(10^{-2.5\cdot(M_{F555W}^{prim}+M_{F555W}^{sec})}})$,
where $M_{F555W}^{Binary}$, $M_{F555W}^{prim}$ $M_{F555W}^{sec}$ are the magnitudes 
of the binary, the primary and the secondary star, respectively.
The latter has been obtained from the isochrone mass/luminosity relation. 
Panel {\sl (b)} in Fig.~\ref{lumfun} shows the comparison between 
the observed and theoretical $\Phi_{norm}$ 
with a binary population.
The inclusion of $\approx$10\% binaries with a flat 
distribution of mass ratios ({\sl q}=0.80) provide a good match 
between theoretical and observed $\Phi_{norm}$ for the models with 
overshooting. A residual discrepancy of $\approx$10\% is 
still present between the observed and  the theoretical $\Phi_{norm}$  as predicted 
by the 
$\Lambda_{os}$=0.00 model.
The adopted binary fraction is 
somewhat smaller than previous estimates ($\le 30$\%) in other LMC and SMC 
clusters \citep{testa95,barmina,chiosi07}.

\section{Discussion and Conclusions}
\label{disc}

The overall CMD characteristics of NGC 1783  are quite similar 
to those of NGC 1978 (Paper I), although there is evidence
of an age difference. Indeed, 
 we have shown that 
the best fit solutions to the observed CMD features  are obtained 
by selecting
$\Lambda_{OS}$=0.1-0.25 and $\tau$=1.2-1.6 Gyr 
for NGC 1783 (see Sect.~\ref{agesec}) and
$\Lambda_{OS}$=0.1 and $\tau$=1.9 Gyr for NGC 1978 (Paper I).

Further insight on the relative age of the two clusters can be obtained from the direct
cluster-to-cluster comparison of the overall CMD properties. To this aim
we can define the $\delta V_{SGB}^{He-Cl}$ parameter  as the magnitude difference     
between the luminosity distribution peak  of the He-Clump and the flat region of
the SGB. This {\it differential} parameter 
can provide an independent estimate of the age, 
and it is formally the analogous of the so-called {\sl vertical method}, 
based on the 
magnitude difference between the TO and the Horizontal Branch magnitude level,
and used to infer the age for the old globulars \citep[see e.g.][]{buo89}.

\begin{figure}[h]
\plotone{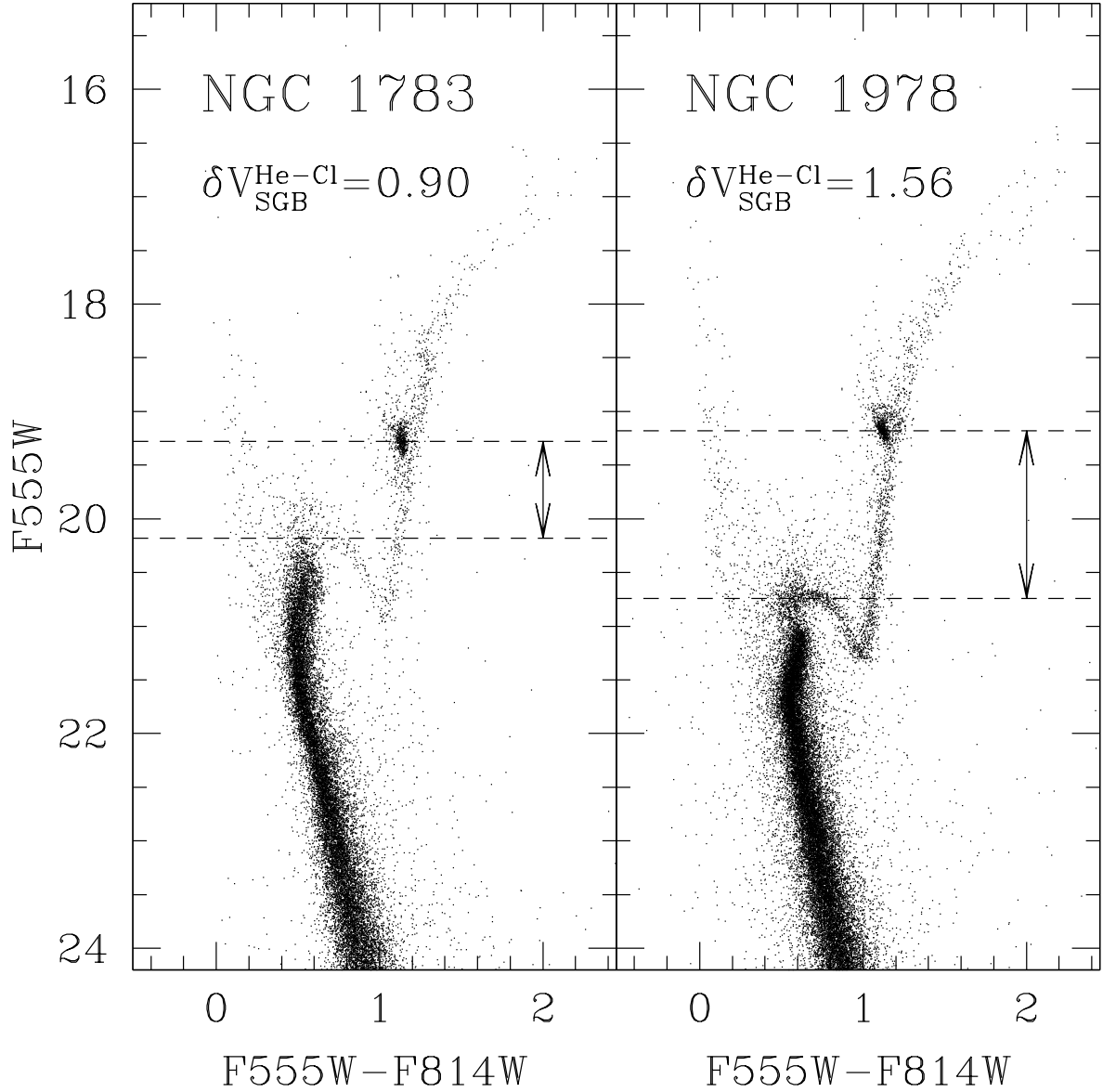}
\caption{ACS@HST (F555W, F555W-F814W) CMDs for the 
LMC cluster NGC 1783 (left panel) and NGC 1978 (right panel). The arrows 
indicate the magnitude difference $\delta V_{SGB}^{He-Cl}$  
between the He-Clump
and the flat portion of the SGB.}
\label{comcmd}
\end{figure}

Fig.~\ref{comcmd} shows the two observed CMDs with  marked the 
$\delta V_{SGB}^{He-Cl}$ parameter:
we find $\delta V_{SGB}^{He-Cl}=0.90$ and $1.56$ for 
NGC1783 and NGC1978, respectively. This difference is an independent,  clearcut
indication  that NGC 1783  is younger than NGC 1978.  

\begin{figure}[h]
\plotone{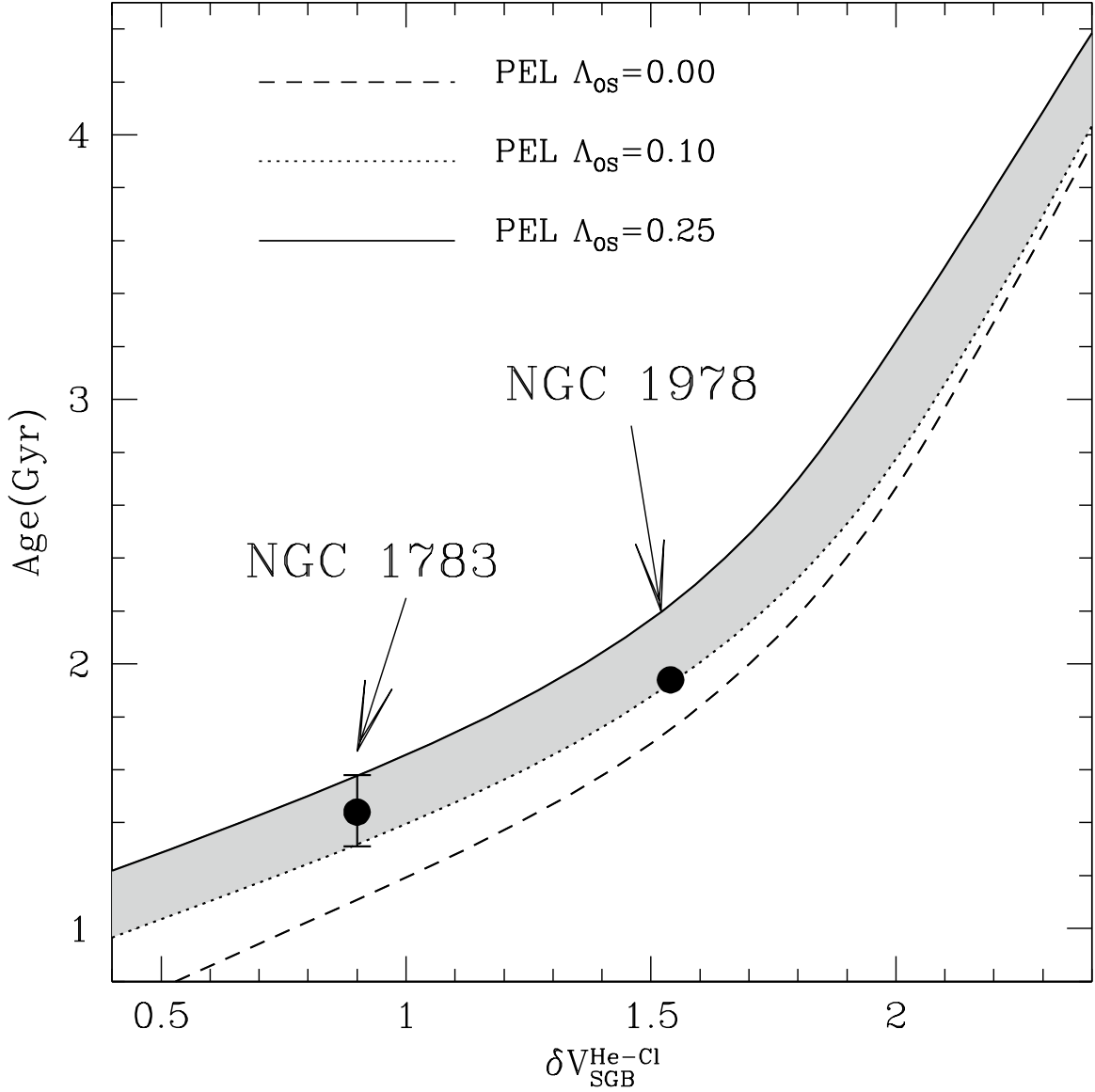}
\caption{Theoretical predictions for the magnitude difference between 
the He-Clump and the flat portion of the SGB as a function of the age for three different 
{\sl overshooting} assumptions: $\Lambda_{OS}$=0.0 (dashed line), $\Lambda_{OS}$=0.10 
(dotted line) and $\Lambda_{OS}$=0.25  (solid line). 
The observed values for $\delta V_{SGB}^{He-Cl}$ and the 
inferred ages for NGC 1783 (this paper) and NGC 1978 (Paper I) are plotted as 
black points.}
\label{theo}
\end{figure}

Fig.~\ref{theo}  shows the
theoretical relations between the $\delta V_{SGB}^{He-Cl}$ observable and the age, 
as derived from the PEL models with different amounts of {\sl overshooting}.
The grey area marked the region of the ($\tau$,
$\delta V_{SGB}^{He-Cl}$) plane for a mild/strong overshooting efficiency
appropriate for NGC 1783.
Hence, by entering the measured $\delta V_{SGB}^{He-Cl}$ in the above relations, 
an independent estimate of the age based on  this differential 
parameter can be obtained. 

By using  the measured value of $\delta V_{SGB}^{He-Cl}=0.90$, 
we find $\tau$=1.4$\pm$0.2$\pm$0.1 Gyr for NGC 1783, 
where the  first errorbar
refers to the uncertainty in {\sl
overshooting} efficiency and the second to the uncertainties in
the adopted reddening and distance modulus.\\ 

This age is still consistent with the
one inferred by  \citet{geisler97} ($\tau$=1.3 Gyr), while it is significantly
older than the age  derived from the s-parameter ($\tau\sim$0.9 Gyr) and 
by  \citet{mould} ($\tau$=0.7-1.1 Gyr).  
\citet{m06} note that
the $N_{Bright-RGB}$/$N_{He-Cl}$ population ratio  computed for NGC 1783   
is too high for the clusters undergoing the RGB Phase-Transition, as
suggested  by the s-parameter age. Our new determination of an older age for
NGC 1783,  better reconcile the $N_{Bright-RGB}$/$N_{He-Cl}$ population ratio
with  the observed well-populated RGB.

Finally, we note that 
the structural parameters ($r_c$, $r_t$) and the age 
of the cluster
inferred from this study, allow us  to
constrain the dynamical state of this cluster.  The resulting core radius of
$r_c$=24.5'' (corresponding to $\sim$5.9 pc adopting the distance modulus of
$(m-M)_0$=18.45, obtained from the best-fit with the {\sl overshooting} models (see
Sect.~\ref{agesec})) is consistent with the age-core radius relationship discussed by
\citet{mac03} and based on the surface brightness radial profiles of 53 LMC rich
clusters. The youngest  (ages $<\sim$200 Myr) clusters of their  sample  exhibit
core radii  $<$3 pc, while the older 
(both intermediate and old-age) 
stellar clusters 
show a more scattered distribution, with $r_c$ between  $\sim$1 and $\sim$8
pc,  a major peak at $r_c\sim$2.5 pc and the presence of several objects with
$r_c>\sim$5 pc. \\ The inferred concentration parameter, c=1.16, is consistent with
a not core-collapse cluster \citep{meylan97}, as expected given the relatively
young age  of NGC 1783.

\acknowledgements  

This research 
was supported by the Agenzia Spaziale Italiana (ASI) and the 
Ministero dell'Istruzione, del\-l'Uni\-versit\`a e della Ricerca.

\newpage

\end{document}